\documentclass[reprint,aps]{revtex4-1}

\usepackage{amsmath}
\usepackage{newtxtext}
\usepackage[slantedGreek,varvw]{newtxmath}

\usepackage{mhchem}

\usepackage{bm}

\usepackage{graphicx}
\usepackage{color}
\usepackage{endnotes}

\newcommand{\ki}{i}
\newcommand{\ke}{e}
\newcommand{\kpi}{\pi}
\newcommand{\dd}{d}
\newcommand{\kk}{k_{\text{B}}}
\newcommand{\tc}{T_{\text{c}}}
\newcommand{\mfc}{\mathfrak{c}}
\newcommand{\sgn}{\operatorname{sgn}}
\newcommand{\Tr}{\operatorname{Tr}}
\newcommand{\tr}{\operatorname{tr}}
\newcommand{\abs}[1]{\lvert#1\rvert}
\newcommand{\average}[1]{\langle #1\rangle}
\newcommand{\daverage}[1]{\langle\!\langle #1\rangle\!\rangle}
\newcommand{\Average}[1]{\left\langle #1\right\rangle}
\newcommand{\hconj}[1]{#1^{\dag}}
\renewcommand{\Im}{\operatorname{Im}}

\let\citen\cite

\begin{document}

%\title{Is the Fock-term negligible within a vortex in weak-coupling superconductors?}
\title{Is the Diagonal Part of the Self-Energy Negligible within an Isolated Vortex in Weak-Coupling Superconductors?}

\author{Noriyuki \surname{Kurosawa}}
\email{kurosawa@vortex.c.u-tokyo.ac.jp}
\affiliation{Department of Basic Science, The University of Tokyo, Meguro, Tokyo 153-8902, Japan}
\date{\today}
\begin{abstract}
  In the weak-coupling theory of superconductivity, the diagonal self-energy term is usually disregarded so that this term is already included in the renormalized chemical potential. Using the bulk solution, we can easily see that the term vanishes in the quasiclassical level. However, the validity of this treatment is obscured in nonuniform systems, such as quantized vortices.
  In this paper, we study an isolated vortex both analytically and numerically using the quasiclassical theory and demonstrate that the finite magnitude of the self-energy can emerge within a vortex in some odd-parity superconductors. We also find that the existence of diagonal self-energy can induce the breaking of the axisymmetry of vortices in chiral $p$-wave superconductors. This implies that the diagonal self-energy is not negligible within a vortex in odd-parity superconductors in general, even in the weak-coupling limit.
\end{abstract}

\maketitle

%%%%%%%%%%%%%%%%%%%%%%%%%%%%%%%%%%%%%%%%%%%%%%%%%%%%%%%%%%%%%%%%%%%%%%%
\section{Introduction}
The strong-coupling theory (or the Eliashberg theory) of superconductivity developed by Migdal\cite{Migdal1958}, Eliashberg\cite{Eliashberg1960}, and Morel and Anderson\cite{Morel1962} is a theory that explicitly deals with the frequencies of the bosons mediating the formation of Cooper pairs. This theory includes the Bardeen--Cooper--Schrieffer (BCS) theory\cite{Bardeen1957} as its weak-coupling limit and is considered to be a more general theory of superconductivity.
In the strong-coupling theory, we deal with the retardation of the effective interaction between electrons. The retardation of the interaction causes the frequency dependence of the modulus of the pair potential in the frequency space, and the diagonal part of the Nambu--Gor'kov space self-energy (hereafter, we call this term just self-energy for brevity), which is the exchange part of the mean-field potential and corresponds to the Fock term, is also frequency-dependent. The gap equations of the Eliashberg theory\cite{Eliashberg1960}, also known as the Eliashberg equations, therefore contain the equation for the self-energy. Once again, we must handle the self-energy explicitly in the strong-coupling theory. Then, how about the weak-coupling theory?

In fact, it is known that the self-energy term shows only a limited effect on a spatially uniform equilibrium system in the weak-coupling limit\cite{Scalapino1969}. Although the chemical potential of the superconducting state is different from that of  the normal state, the shift is on the order of \(\Delta^2/\mu_{\text{F}}\), where \(\Delta\) is the magnitude of the pair potential, and \(\mu_{\text{F}}\) is the Fermi energy. For many superconductors, this difference is very small and can be negligible, except for some unconventional superconductors.
Thus, we consider that the self-energy is included in the definition of the chemical potential and is usually dropped from the equations of the weak-coupling theory of superconductivity. 

On the other hand, however, when we consider spatially nonuniform systems, the validity of this treatment becomes unclear. For example, within a vortex, there are low-energy states that do not exist in the bulk system\cite{Caroli1964}. The effect of these states on the self-energy is not necessarily the same as those of bulk states.
The low-energy bound states within the vortex dominate the physics of type-II superconductors in the vortex phase. For example, it is indispensable to consider these bound states when discussing the flow conductivity of vortices\cite{Kopnin1976,Kopnin1991,Kopnin1997}. Thus, the study of the low-energy states in nonuniform systems is significant not only for theoretical interests but also for applications of superconductivity.
Nevertheless, the treatment of the term in previous studies has been insufficient. There have been studies of superfluid helium-3\cite{Fogelstroem1995,Silaev2015}. However, for electronic superconductors, as far as we know, current status is no more than connecting the bulk solutions between normal and superconducting states\cite{McMillan1968}.
There is no clear evidence that such a treatment gives a good description of the nonuniform systems, such as vortices. Besides, when we analyze and discuss experimental results, it is desirable to have a detailed knowledge about each low-energy quasiparticle, which is inaccessible via the above simple theory. For these reasons, we study the effect of the diagonal self-energy term, to which little attention has been paid so far, on a vortex.

As an earlier short report, we published a conference proceeding\cite{Kurosawa2017}, in which we focused on the strong-coupling chiral $p$-wave superconductors, for which we cannot neglect the self-energy apparently due to its frequency dependence of the self-energies. 
In this paper, we consider more general chiral superconductors [specified by the d-vector in Eq.~\eqref{eq:d-vector of chiral superconductor}]. { However, we concentrate on the weak-coupling regime (i.e., there is no frequency dependencie of pair potentials and self-energies) to understand the effect of this term more clearly.} We show that the phenomenon reported in Ref.~\citen{Kurosawa2017} can be regarded as an effect of the self-energy, rather than an effect of the frequency dependence on the pair potential.

This paper is organized as follows.
In Sect.~2, we present the model and formulation used in this study.
In Sect.~3, we study the self-energy in the axisymmetric vortices in chiral superconductors by an analytic method. We also confirm the obtained result by numerical calculations (we present the details in Appendix~\ref{appendix:numerical}). In Sect.~4, we show that the non-axisymmetric vortices are stable in chiral $p$-wave superconductors owing to the existence of the self-energy term. Finally, Sect.~5 summarizes our findings and gives concluding remarks.

%%%%%%%%%%%%%%%%%%%%%%%%%%%%%%%%%%%%%%%%%%%%%%%%%%%%%%%%%%%%%%%%%%%%%%%
\section{Formulation}\label{sec:formulation}
In this paper, we use the quasiclassical theory developed by Eilenberger\cite{Eilenberger1968} and Larkin and Ovchinnikov\cite{Larkin1969}. This method is suited to a superconducting system whose characteristic length of the pair potential \(\xi_0 = \hbar v_{\text{F}}/(\kk \tc)\), where \(v_{\text{F}}\) is the Fermi velocity and \(\tc\) is the transition temperature of superconductivity, is much larger than the Fermi wavelength \(k_{\text{F}}^{-1}\).
In this theory, we target the quasiclassical Green's function \(\check g(z,\hat k,\bm{r})\), which we define in Appendix~\ref{appendix:quasiclassical}. We use \(\hat k=\bm{k}/\abs{k}\) for the direction of the quasiparticle's momentum, $\bm{r}$ for the position, and \(z\) to denote both fermionic Matsubara frequencies \(\ki\epsilon_n = \ki\kpi(2n+1)k_{\text{B}}T/\hbar\) and real frequencies. The Green's function $\check g(z,\hat k,\bm{r})$ is a $4\times 4$ matrix of Nambu--Gor'kov space and spin space, and we also write it as
\begin{align}
\check g(z,\hat k,\bm{r}) &= \begin{pmatrix}
g(z,\hat k,\bm{r}) & f(z,\hat k,\bm{r}) \\ -\bar f(z,\hat k,\bm{r}) & -g(z,\hat k,\bm{r})
\end{pmatrix}
.
\end{align}

In this paper, we focus on the chiral superconductors, whose d-vector $\bm{d}$ has the form
\begin{align}
 \bm{d} = {\hat k_z}^{l-m}(\hat k_x\pm\ki \hat k_y)^m\hat z
 \label{eq:d-vector of chiral superconductor},
\end{align}
where $l$ and $m$ are natural numbers and $0\le m \le l$.
When we consider singlet superconductors, we can treat \(g\), \(f\), and \(\bar f\) as scalar (complex) values instead of \(2\times 2\) matrices. In triplet cases, we must deal with them as matrices in general. However, if the d-vector has the form of Eq.~\eqref{eq:d-vector of chiral superconductor}, they can be reduced to scalars even in triplet cases. Hereafter, we consider $\check g(z,\hat k,\bm{r})$ to be a $2\times 2$ matrix of complex numbers.

If we take the weak-coupling limit of the Eliashberg equations\cite{Eliashberg1960,Scalapino1969} and its quasiclassical form, we can obtain the following gap equations:
\begin{subequations}
\begin{align}
\Sigma(\hat k,\bm{r}) &= \kk T\sum_{n}^{\abs{\epsilon_n}<\epsilon_{\text{c}}} \average{N_0 v(\hat k,\hat k') g(\ki\epsilon_n,\hat k',\bm{r}) }_{\hat k'}
\label{eq:weak-coupling-eliashberg-equation-diagonal}
,
\\
\Delta(\hat k,\bm{r}) &= \kk T\sum_{n}^{\abs{\epsilon_n}<\epsilon_{\text{c}}} \average{N_0 [v(\hat k,\hat k')-\mu^{*}] f(\ki\epsilon_n,\hat k',\bm{r}) }_{\hat k'}
\label{eq:weak-coupling-eliashberg-equation-off-diagonal}
,
\end{align}%
\label{eq:weak-coupling-eliashberg-equation}%
\end{subequations}
where \(v(\hat k,\hat k')\) is the effective interaction between two electrons, \(\epsilon_{\text{c}}\) is the cutoff of the Matsubara frequency, \(N_0\) is the density of states upon the Fermi level, and \(\Sigma\) is the self-energy that corresponds to the Fock term. For simplicity, we assume that the Fermi surface is isotropic. The symbol \(\mu^{*}\) is the pseudo Coulomb potential part of the effective electron-electron interaction. We regard \(\mu^{*}\) as negligibly small for a while. Equation \eqref{eq:weak-coupling-eliashberg-equation-diagonal} does not contain \(\mu^{*}\) because this term is considered to be already included in the chemical potential.
The notation \(\average{\dots}_{\hat k}\) denotes an average over the Fermi surface. For two-dimensional(2D) systems with circular Fermi surface,
\begin{align}
\hat k &= (\cos\phi,\sin\phi)
,\nonumber\\
\average{ X(\hat k) }_{\hat k} &= \int_0^{2\kpi}\frac{\dd\phi}{2\kpi} X(\phi)
,
\end{align}
and for three-dimensional (3D) systems with a spherical Fermi surface,
\begin{align}
\hat k &= (\sin\theta\cos\phi,\sin\theta\sin\phi,\cos\theta)
,\nonumber\\
\average{ X(\hat k) }_{\hat k} &= \int_0^\pi\dd\theta\int_0^{2\kpi}\dd\phi\frac{\sin\theta}{4\kpi} X(\theta,\phi)
.
\end{align}
In the quasiclassical theory, we take the chemical potential as the origin of the energy level. This immediately yields \(\Sigma=0\) in the bulk from the definition of the chemical potential.

Since we assume that the system is isotropic in the momentum space, the effective coupling function \(v(\hat k,\hat k')\) can be decoupled as
\begin{align}
v(\phi,\phi') &= \sum_{m=-\infty}^\infty c_{m}\ke^{\ki m\phi}\ke^{-\ki m\phi'}
\end{align}
in 2D systems and
\begin{align}
v(\theta,\phi,\theta',\phi') &= 4\kpi\sum_{l=0}^\infty\sum_{m=-l}^l c_{l,m}Y_{l,m}(\theta,\phi)Y^{*}_{l,m}(\theta',\phi')
\end{align}
in 3D systems.
The function \(Y_{l,m}(\theta,\phi)\) is the spherical harmonic function, which is defined using the associated Legendre function \(P_{l,m}(\cos\theta)\) as
\begin{align}
Y_{l,m}(\theta,\phi) &= \sqrt{\frac{(2l+1)}{4\kpi}\frac{(l-\abs{m})!}{(l+\abs{m})!}} P_{l,m}(\cos\theta)\ke^{\ki m\phi}
.
\end{align}

The quasiclassical Green's function \(\check g\) obeys the Eilenberger equation\cite{Eilenberger1968}
\begin{align}
\ki \hbar\bm{v}_{\text{F}}\cdot\nabla\check g + [\hbar z\check\tau_3+q\bm{v}_{\text{F}}\cdot\bm{A}\check\tau_3-\check\Sigma,\;\check g] &= 0
\label{eq:eilenberger-equation}
\end{align}
and satisfies the normalization condition \(\check g^2 = -\kpi^2\check\tau_0\),
where \(\bm{A}(\bm{r})\) is the vector potential, \(q\) is the charge of the fermion, \(\check\tau_i\) ($i=0,1,2,3$) are the unit matrix and the Pauli matrices, and
\begin{align}
\check \Sigma &= \begin{pmatrix} \Sigma & \Delta \\ -\Delta^{*} &-\Sigma \end{pmatrix}
\end{align}
is the self-energy in Nambu--Gor'kov space.
The bulk solution to Eq.~\eqref{eq:eilenberger-equation} is
\begin{align}
\check g(z,\hat k) &= \frac{\kpi}{\sqrt{-(\hbar z-\Sigma)^2+\abs{\Delta}^2}}
\begin{pmatrix}
-\hbar z+\Sigma & \Delta \\ -\Delta^{*} & +\hbar z-\Sigma
\end{pmatrix}
,
\label{eq:bulk-quasiclassical-greens-function}
\end{align}
and from Eqs.~\eqref{eq:weak-coupling-eliashberg-equation-diagonal} and \eqref{eq:bulk-quasiclassical-greens-function}, we can confirm that \(\Sigma = 0\) in uniform systems again.

Once we have obtained \(\check g\), \(\check \Sigma\), and \(\bm{A}\) self-consistently, we can also obtain the free energy \(\mathcal{J}\) as\cite{Eliashberg1963,Bardeen1964,Serene1983,Thuneberg1984}
\begin{align}
\mathcal{J}-\mathcal{J}_{\text{n}} &= \Tr_2 N_0\kk T\sum_n\int\dd\bm{r}\left[\int_0^1\dd s\average{\check g_s\check\Sigma}-\frac{1}{2}\average{\check g\check\Sigma}\right]
\nonumber\\
&\qquad+\int\dd\bm{r}\left[\frac{1}{2\mu_0}(\bm{B}^2-\bm{B}_{\text{n}}^2)\right]
,
\end{align}
where \(\Tr_2\) is the trace of the Nambu--Gor'kov space, \(\mu_0\) is the magnetic constant, \(\bm{B} = \nabla\times\bm{A}\) is the magnetic field, \(\mathcal{J}_{\text{n}}\) is the free energy of the normal state, \(\bm{B}_{\text{n}}\) is the magnetic field of the normal state, and \(\check g_s\) is the solution of
\begin{align}
\ki \hbar\bm{v}_{\text{F}}\cdot\nabla\check g_{s} + [\hbar z\check\tau_3+q\bm{v}_{\text{F}}\cdot\bm{A}\check\tau_3-s\check\Sigma,\;\check g_{s}] &= 0
.
\end{align}
We only need the difference in the free energy between various vortices; accordingly, we ignore the term \(\bm{B}_{\text{n}}^2\).
The vector potential \(\bm{A}\) is obtained by the Maxwell equation (Ampère's law):
\begin{align}
\nabla\times[\nabla\times\bm{A}(\bm{r})] &= \mu_0\bm{j}(\bm{r})
,
\end{align}
where \(\bm{j}\) is the electric current density calculated as
\begin{align}
\bm{j}(\bm{r}) &= 2q\kk T N_0\sum_{\epsilon_n}\average{\bm{v}_{\text{F}}(\hat k)g(\ki\epsilon_n,\hat k,\bm{r})}_{\hat k}
.
\end{align}
We also define the characteristic length of the magnetic penetration \(\lambda_{\text{M}}\) as
\begin{align}
(\lambda_{\text{M}})^{-2} &= \mu_0v_{\text{F}}^2q^2N_0,
\end{align}
which is the London penetration depth up to a numerical factor.
In the calculation carried out in the following sections, we set \(\lambda_{\text{M}}/\xi_0 = 2.5\).

In this paper, we focus on an isolated vortex. When we analytically deal with the vortex, we assume that the vortex is axisymmetric and the pair potential \(\Delta\) around the vortex has the form
\begin{align}
\Delta(\phi,\bm{r}) &= \Delta_0(r)\ke^{\ki m\phi}\ke^{+\ki \varphi}
,\\
\Delta(\theta,\phi,\bm{r}) &= \Delta_0(r)Y_{l,m}(\theta,\phi)\ke^{+\ki \varphi}
\end{align}
for simplicity, where \(\bm{r} = (r\cos\varphi,r\sin\varphi)\), \(\Delta_0(r)\) is a non-negative real-valued function, and \(\Delta_0(r)\simeq \Delta_\infty r/\xi_1\) for \(r\ll \xi_0\) and \(\xi_1 > 0\). We use \(\Delta_{\infty} = \Delta_0(r\to\infty)\) as the magnitude of the bulk pair potential. Although there are also induced components of the pair potential, we ignore them in the following analytical discussions for simplicity.
\begin{figure}
\centering
\includegraphics[width=0.6\columnwidth]{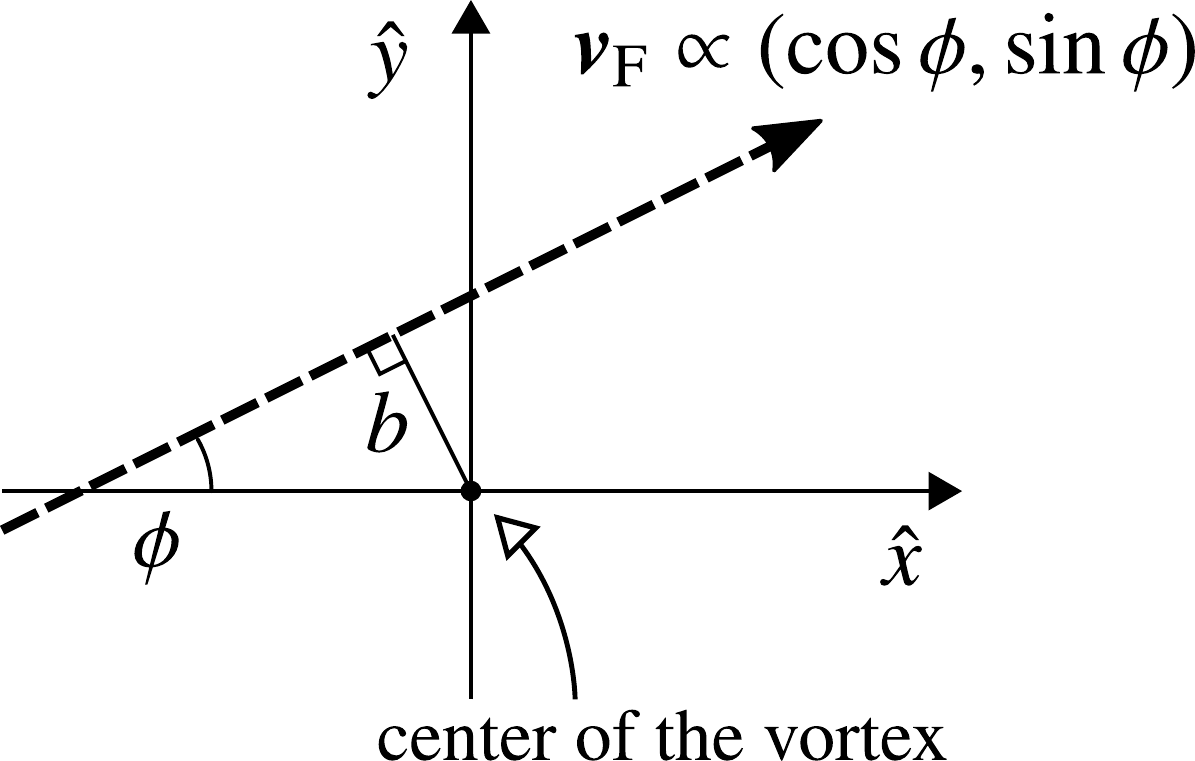}
\caption{Impact parameter $b$ and the coordinate system. The dashed line denotes the trajectory of the quasiparticle with momentum $\phi$ and impact parameter $b$.}
\label{fig:coordinate-system}
\end{figure}
In this case, the quasiclassical Green's function near the vortex core can be obtained by the perturbative method developed by Kramer and Pesch\cite{Kramer1974} (hereafter we call this method the \textit{Kramer--Pesch approximation}).
The expansion parameter of this method is \(b=r\sin(\varphi-\phi)\), which corresponds to the impact parameter of quasiclassical quasiparticles (see Fig.~\ref{fig:coordinate-system}).
The solution of the quasiclassical Green's function by the Kramer--Pesch approximation\cite{Kramer1974,Kato2000,Nagai2006} is
\begin{align}
\check g(z,\theta,\phi,r,\varphi) &\simeq \frac{2\kpi\ke^{-u^\theta(r)}/C^\theta}{z-E^\theta b-\tilde\Sigma/\hbar}\check M^\theta
,
\label{eq:green-function-kramer-pesch}
\end{align}
where 
\begin{subequations}
\begin{align}
u^\theta(s) &= \frac{2}{\hbar v_{\text{F}}^\theta}\int_0^{\abs{s}}\dd s' \Delta_0^\theta(s')
,\displaybreak[1]\\
C^\theta &= \frac{4}{v_{\text{F}}^\theta}\int_{0}^{s_{\text{c}}}\dd s'\ke^{-u^\theta(s')}
,\displaybreak[1]\\
E^\theta &= \frac{4}{\hbar v_{\text{F}}^\theta C^\theta}\int_0^{s_{\text{c}}}\dd s'\frac{\Delta_0^\theta(s')}{s'}\ke^{-u^\theta(s')}
,\displaybreak[1]\\
\tilde\Sigma(z,\theta,\phi,b) &= \frac{2}{v_{\text{F}}^\theta C^\theta}\int_{-s_{\text{c}}}^{s_{\text{c}}}\dd s'\Sigma(z,\theta,\phi,s',b)\ke^{-u^\theta(s')}
,\displaybreak[1]\\
\check M^\theta &= \begin{pmatrix} 1 & -\ki\varsigma^\theta\ke^{+\ki(m+1)\phi} \\ -\ki\varsigma^\theta\ke^{-\ki(m+1)\phi}  & -1 \end{pmatrix}
\label{eq:definition-of-Sigma-tilde}
,
\end{align}
\end{subequations}
where \(\Delta_0^\theta(r) = \Delta_0(r)\) (2D case) or \(\Delta_0(r)\abs{Y_{l,m}(\theta,\phi)}\) (3D case), 
\(v_{\text{F}}^\theta = v_{\text{F}}\) (2D case) or \(v_{\text{F}}\sin\theta\) (3D case), \(\varsigma^\theta=1\) (2D case) or \(\sgn [Y_{l,m}(\theta,\phi)\ke^{-\ki m\phi}]\) (3D case),
\(\Sigma(z,\theta,\phi,s,b) = \Sigma(z,\theta,\phi,r=\sqrt{s^2+b^2},\varphi=\arg(s+\ki b)+\phi)\), and \(s_{\text{c}} \gg \xi_0\) is the cutoff length. 

\begin{figure}
\centering
\includegraphics[width=0.475\columnwidth]{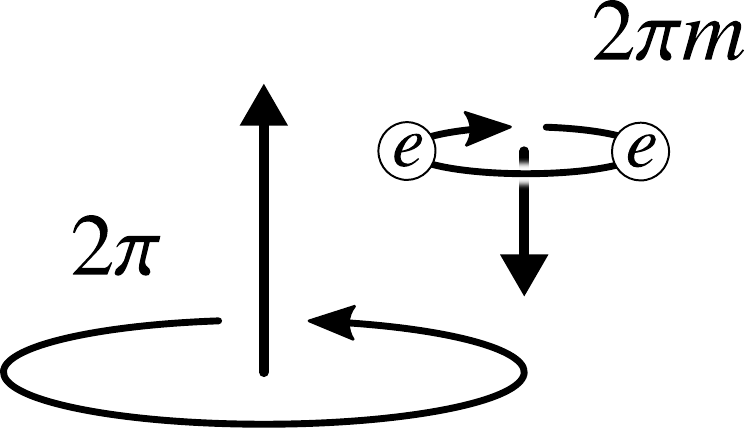}
\hspace{0.5em}
\includegraphics[width=0.475\columnwidth]{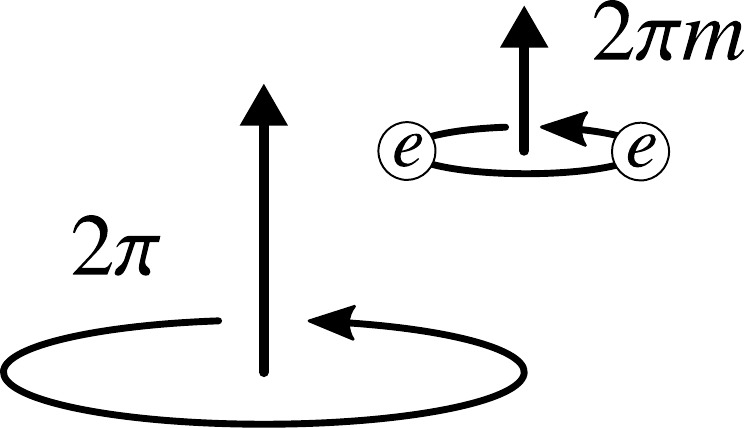}
\caption{Schematic pictures of an \textit{antiparallel vortex} ($m<0$) and a \textit{parallel vortex} ($m>0$). Left: \textit{antiparallel}, right: \textit{parallel}.}
\label{fig:schematic-parallel-antiparallel}
\end{figure}

The order parameter corresponding to the d-vector [Eq.~\eqref{eq:d-vector of chiral superconductor}] has the form \(\Delta(\theta,\phi,\bm{r}) = \Delta(\theta,\bm{r})\ke^{\pm\ki m\phi}\), where \(l=1\) for a chiral $p$-wave superconductor, \(l=2\) for a chiral $d$-wave superconductor, and \(l=3\) for a chiral $f$-wave superconductor.
There are two types of vortices in the chiral superconductors\cite{Kato2000,Matsumoto1999,Hayashi2005,Tanuma2009,Sauls2009}; one type of vortex has the angular momentum (vorticity) parallel to the angular momentum of the Cooper pairs (chirality), and the other type has the antiparallel one. The pair potential of the former vortex has the form \(\Delta_0(r)\ke^{+\ki m\phi + \ki\varphi}\) and the latter has the form \(\Delta_0(r)\ke^{-\ki m\phi + \ki\varphi}\) for positive \(m\) far from the vortex core. In this paper, we call them the \textit{parallel vortex} and the \textit{antiparallel vortex}, respectively (Fig.~\ref{fig:schematic-parallel-antiparallel}).
Because the momentum or wavelength is not a good quantum number around a vortex, there exists an induced component that has opposite chirality against the bulk (e.g., see Ref.~\citen{Sauls2009}). The vorticity of the induced component is determined so that the total angular momentum is the same as the dominant one. For example, if the major component of the pair potential is \(\Delta_{+}(r)\ke^{+\ki m\phi + \ki\varphi}\), the minor component has the form \(\Delta_{-}(r)\ke^{-\ki m\phi + \ki(2m+1)\varphi}\).

%%%%%%%%%%%%%%%%%%%%%%%%%%%%%%%%%%%%%%%%%%%%%%%%%%%%%%%%%%%%%%%%%%%%%%%
\section{Emergence of Self-Energy in Vortices}

Using the solution of the Kramer--Pesch approximation for \(\check g\) around a vortex core, we can discuss the self-energy term within a vortex.
Firstly, we consider the 2D case.
Substituting Eq.~\eqref{eq:green-function-kramer-pesch} into the gap equation \eqref{eq:weak-coupling-eliashberg-equation-diagonal}, the \(m\)-mode of the self-energy \(\Sigma_{m}\) is
\begin{align}
\Sigma_{m}(\theta,\phi,\bm{r}) 
&= c_{m} N_0k_{\text{B}}T\sum_n \Average{\ke^{\ki m\phi-\ki m\phi'}g(\ki\epsilon_n,\theta,\phi',r,\varphi)}_{\hat k'}
\nonumber\\
&\simeq
D_{m}^{(2)}(\theta,r)\ke^{\ki m\phi}\sum_n\Average{\frac{\ke^{-\ki m\phi'}}{\ki\epsilon_n-E^\theta r\sin(\varphi-\phi')}}_{\hat k'}
\nonumber\\
&=
D_{m}^{(2)}(\theta,r)\ke^{\ki m\phi-\ki m\varphi}\sum_n\Average{\frac{\ke^{-\ki m\phi'}}{\ki\epsilon_n+E^\theta r\sin\phi'}}_{\hat k'}
\nonumber\\
&=D_{m}^{(2)}(\theta,r)\ke^{\ki m(\phi-\varphi)}\sum_{n\ge0}\Average{\frac{2E^\theta r\sin\phi'\ke^{-\ki m\phi'}}{\epsilon_n^2+(E^\theta r\sin\phi')^2}}_{\hat k'}
\label{eq:sigma_lm_2d_simple}
\end{align}
with \(\hat k' = (\cos\phi', \sin\phi') \), where
\begin{align}
D_{m}^{(2)}(\theta,r) &= \frac{2\kpi k_{\text{B}}T N_0c_{m}\ke^{-u^\theta(r)}}{C^\theta}
,
\end{align}
and we ignore \(\tilde\Sigma\) in \(g\) in Eq.~\eqref{eq:sigma_lm_2d_simple} for a short while.
At sufficiently near the vortex core such that \(E^\theta r < \kpi \kk T/\hbar\), Eq.~\eqref{eq:sigma_lm_2d_simple} yields
\begin{align}
\eqref{eq:sigma_lm_2d_simple}
&=
D_{m}^{(2)}(\theta, r)\ke^{\ki m(\phi-\varphi)}\sum_{n\ge0}\frac{-2\ki\sgn(m)(1-\alpha_{-}^2)\alpha_{-}^{\abs{m}}}{E^\theta r\alpha_{-}(\alpha_{+}^2-\alpha_{-}^2)}
\nonumber\\
&\qquad\times[1-(-1)^{\abs{m}}]
\label{eq:sigma_lm_2d_odd-even}
,
\end{align}
where
\begin{align}
\alpha_{\pm} &= \left[1+2\epsilon_n^2/(E^\theta r)^2\pm2\sqrt{\epsilon_n^2/(E^\theta r)^2+\epsilon_n^4/(E^\theta r)^4}\right]^{1/2}
.
\end{align}
From Eq.~\eqref{eq:sigma_lm_2d_odd-even}, we can see that \(\Sigma_{l,m}\simeq 0\) for even \(m\) and \(\Sigma_{l,m}\ne 0\) for odd \(m\). When \(\abs{m}=2m'+1\) and \(E^\theta r\ll \kpi\kk T/\hbar\), the approximate self-energy is 
\begin{align}
\Sigma_{m}(\theta,\phi,\bm{r})
&\simeq
D_{m}^{(2)}(\theta,r)\ke^{\ki m(\phi-\varphi)}\sum_{n\ge 0} (\ki\sgn(m) E^\theta r)\frac{(\alpha_{-})^{m'}}{\epsilon_n^2}
\nonumber\\
&\simeq
\ki\sgn(m)\frac{(E^\theta r)^{m'+1}}{2^{m'}}D_{m}^{(2)}(\theta,r)\ke^{\ki m(\phi-\varphi)}\sum_{n\ge 0}\frac{1}{\epsilon_n^{m'+2}}
\label{eq:2d-kramer-pesch-sigma-lm}
,
\end{align}
and we can see that \(\Sigma_{m}(\bm{r}) \propto r\) for \(m=\pm 1\) and \(\Sigma_{m}(\bm{r}) \propto r^2\) for \(m = \pm3\).

For 3D cases, the integration of \(\exp[\pm\ki m\phi]\) with respect to \(\phi\) is completely same as that of the 2D one; within the range \(l\le 3\), only the terms \((l,m)=(1,\pm 1)\), \((2,\pm 1)\), \((3,\pm 1)\), \((3,\pm 3)\) survive after the integration with respect to \(\phi\).
In addition, we must also consider the integration with respect to \(\theta\) in 3D cases. Because $u^\theta=u^{\kpi-\theta}$, $C^\theta=C^{\kpi-\theta}$, and $E^\theta=E^{\kpi-\theta}$, we can see that \(g(\ki\epsilon_n,\theta,\phi,\bm{r})\simeq g(\ki\epsilon_n,\kpi-\theta,\phi,\bm{r})\). This yields 
\begin{align}
\int_0^{\kpi}\dd\theta\sin\theta P_{l,m}(\cos\theta)g(\ki\epsilon_n,\theta,\phi,r,\varphi) &= 0
\end{align}
for \((l,m)=(2,1)\) in the vicinity of the core.
Therefore, we conclude that the \((l,m)\) mode of the self-energy \(\Sigma_{l,m}\) can be induced within a vortex in an odd-parity superconductor such that \((l,m)=(1,\pm1)\), \((3,\pm1)\), \((3,\pm3)\).

\begin{figure}
\centering
\includegraphics[width=.48\columnwidth]{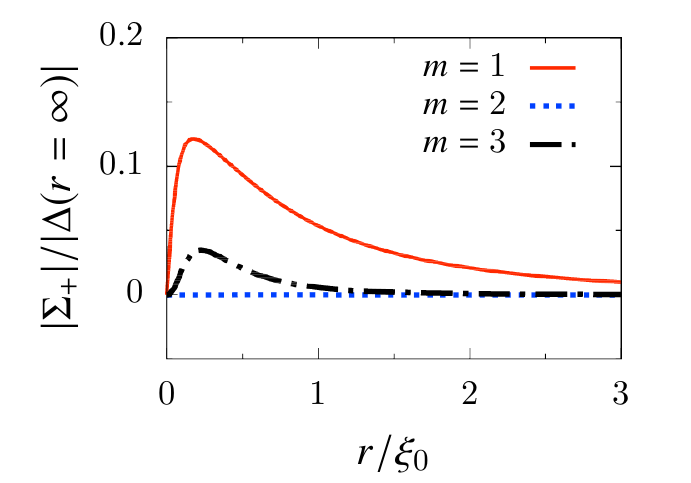}
\includegraphics[width=.48\columnwidth]{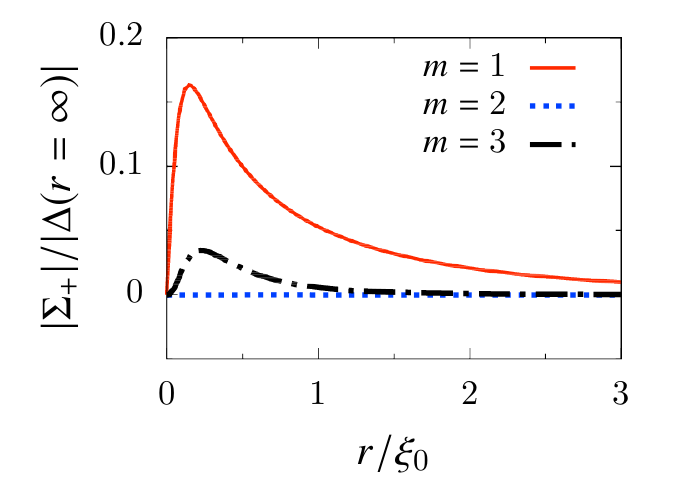}
\caption{(Color online) Self-energies of isolated axisymmetric vortices in 2D chiral superconductors at $T/\tc=0.2$. Left: antiparallel vortices, right: parallel vortices.}
\label{fig:2d-axisymmetrix-fock-term}
\end{figure}

\begin{figure}
\centering
\includegraphics[width=.48\columnwidth]{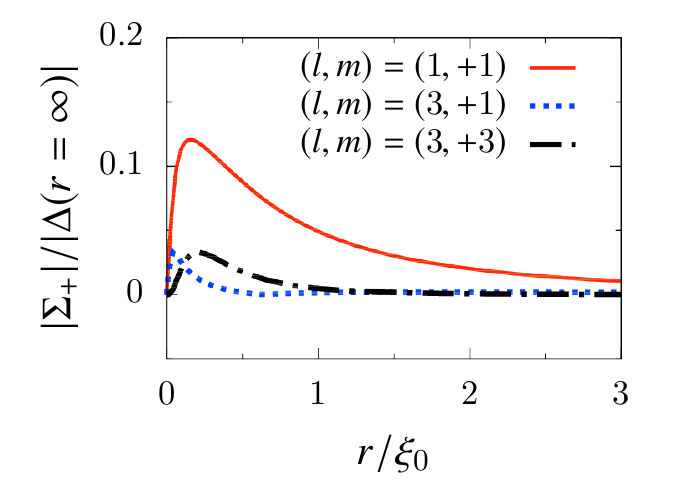}
\includegraphics[width=.48\columnwidth]{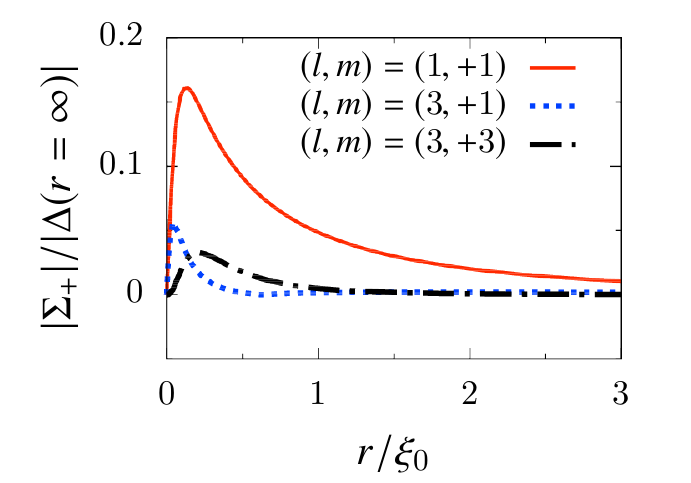}
\caption{(Color online) Self-energies of isolated axisymmetric vortices in 3D chiral superconductors at $T/\tc=0.2$. Self-energies with $(l,m)=(2,1)$, $(2,2)$, $(3,2)$ disappear. Left: antiparallel vortices, right: parallel vortices.}
\label{fig:3d-axisymmetrix-fock-term}
\end{figure}

Figures \ref{fig:2d-axisymmetrix-fock-term} and \ref{fig:3d-axisymmetrix-fock-term} show the results of the self-consistent calculations of vortices in chiral superconductors. We can confirm the above discussion via the numerical calculation: the self-energy arises in a vortex of odd-parity superconductors (odd $l$) with odd $m$, and disappears otherwise.

We can also see that the \(s\)-wave component vanishes within the vortex. Therefore, the existence of the finite magnitude of the pseudo Coulomb potential \(\mu^{*}\) does not modify the results and discussions in this section at all, as long as \(\mu^{*}\) is isotropic.

Next, we consider the effect of the self-energy on the Green's function via \(\tilde\Sigma\) [or its \(((l,)m)\)-mode \(\tilde\Sigma_{(l,)m}\)].
Equation \eqref{eq:2d-kramer-pesch-sigma-lm} yields \(\Sigma_{(l,)m=\pm1} \propto b\) and \(\Sigma_{(l,)m=\pm3} \propto b^2\) and therefore,
\begin{align}
\tilde\Sigma_{(l,)m} &\propto
\begin{cases}
b &\text{for }m=1,
\\
b^2 &\text{for }m=3.
\end{cases}
\label{eq:tilde-sigma-b-dependence}
\end{align}
Because the expansion parameter of the Kramer--Pesch approximation is \(b\), \(\tilde\Sigma\) is regarded as a small correction to \(E^\theta\) in Eq.~\eqref{eq:green-function-kramer-pesch}. 
Furthermore, because
\begin{subequations}
\begin{align}
\int_0^{\kpi}\dd\theta\sin\theta P_{1,1}(\cos\theta) &= \frac{\sqrt{3}\kpi}{2},
\\
\int_0^{\kpi}\dd\theta\sin\theta P_{3,1}(\cos\theta) &= \frac{3\kpi}{16},
\end{align}
\end{subequations}
we can expect that the diagonal self-energy term in a vortex of (\(l=3\), \(m=1\)) superconductivity is fairly smaller than that of (\(l=1\), \(m=1\)). From Fig.~\ref{fig:3d-axisymmetrix-fock-term}, we can confirm the above discussions. Our findings in this section are summarized as follows: the self-energy \(\Sigma\) is finite but small when \(m=\pm3\) or \(l\ge 2\), and the induced \(\Sigma\) around a vortex is not so small but the effect on the Green's function is renormalized into the dispersion relation of quasiparticles \(E^\theta\) when \(l=1\) and \(m=\pm1\), as long as the Kramer--Pesch approximation is valid.

%%%%%%%%%%%%%%%%%%%%%%%%%%%%%%%%%%%%%%%%%%%%%%%%%%%%%%%%%%%%%%%%%%%%%%%
\section{Emergence of Non-Axisymmetric Vortices}

In the last part of the previous section, we saw that we can renormalize the self-energy term into the dispersion of the quasiparticles. 
%This seems to indicate that the existence of self-energy has no significant effect on the vortices. Contrary to the intuition, however, we find that the self-energy term affects vortices qualitatively as well as quantitatively.
We present a more clear and profound influence on the \textit{parallel vortices} (see the last paragraph of Sect.~\ref{sec:formulation} for the definition) in chiral $p$-wave superconductors.

\begin{figure}
\centering\includegraphics[width=.96\columnwidth]{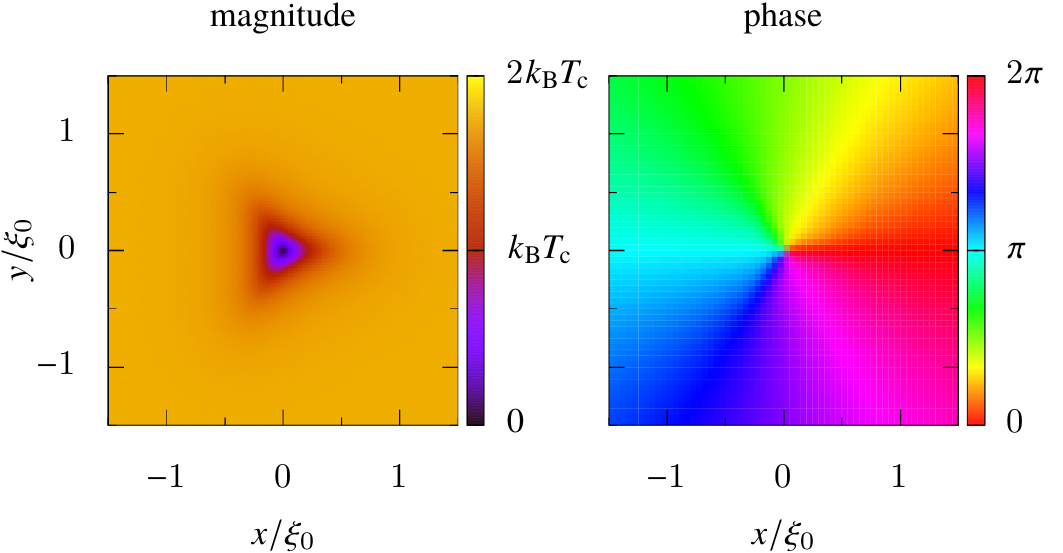}
\caption{(Color online) Major component (\(\Delta_{+}\)) of the non-axisymmetric triangular parallel vortex in 2D chiral $p$-wave superconductors at \(T/\tc=0.2\).}
\label{fig:non-axisymmetric-vortex-major}
\end{figure}

\begin{figure}
\centering
\includegraphics[width=.96\columnwidth]{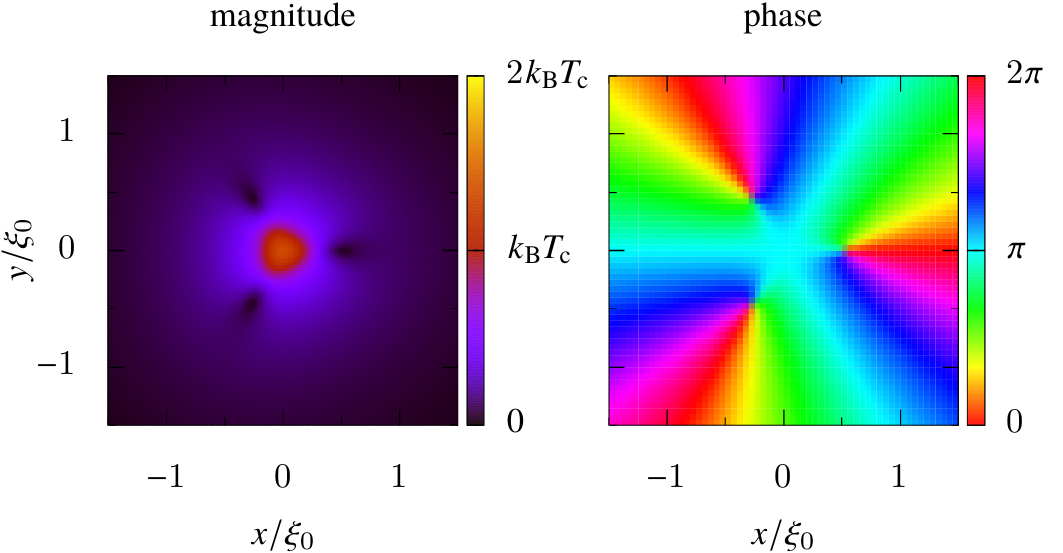}
\caption{(Color online) Minor component (\(\Delta_{-}\)) of the non-axisymmetric triangular parallel vortex in 2D chiral $p$-wave superconductors at \(T/\tc=0.2\).}
\label{fig:non-axisymmetric-vortex-minor}
\end{figure}

\begin{figure}
\centering
\includegraphics[width=.96\columnwidth]{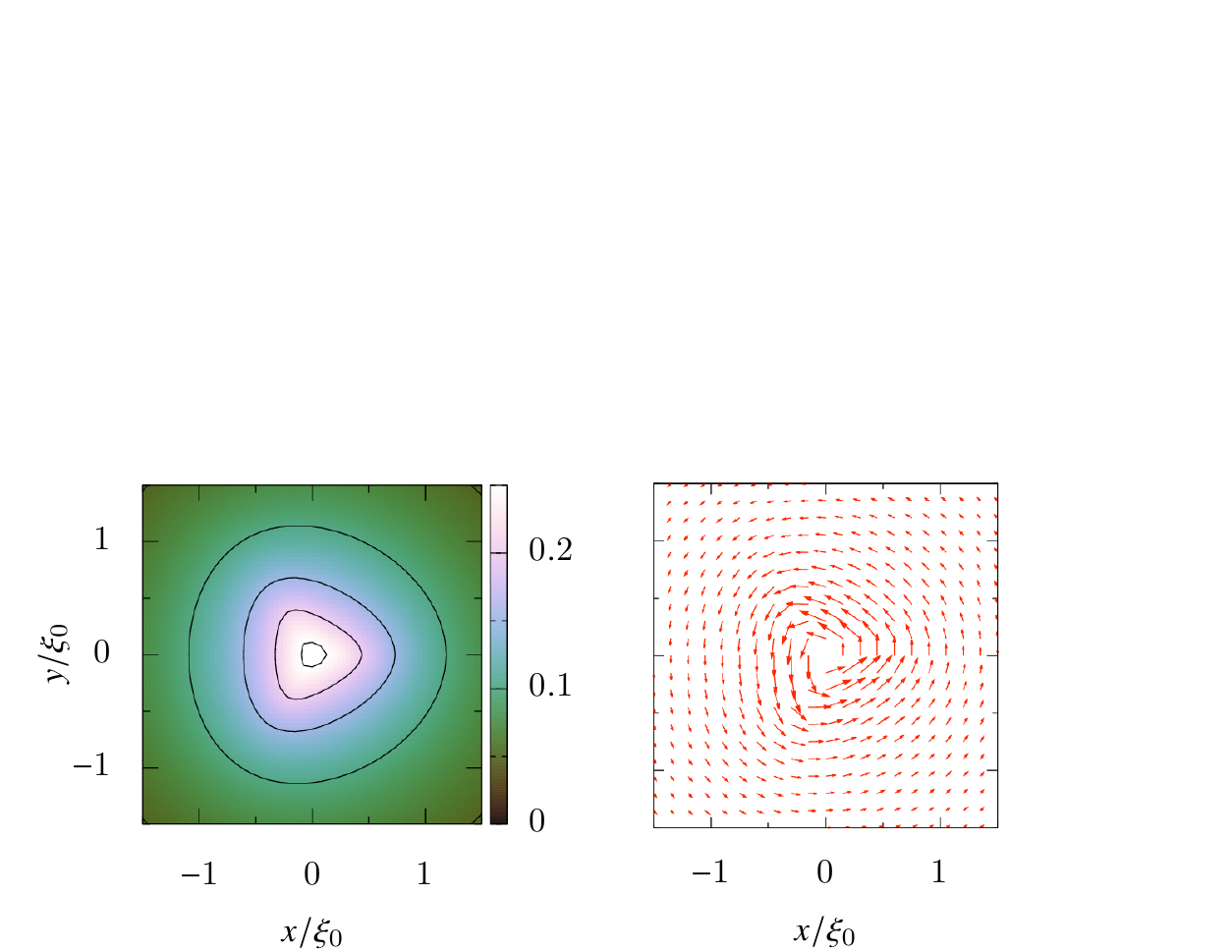}
\caption{(Color online) Magnetic field $qB/(\hbar/\xi_0^2)$ (left) and electric current density (right) of the non-axisymmetric triangular parallel vortex in 2D chiral $p$-wave superconductors at \(T/\tc=0.2\).}
\label{fig:nonaxisymmetric-electromagnetic}
\end{figure}

\begin{figure}
\centering
\includegraphics[width=.96\columnwidth]{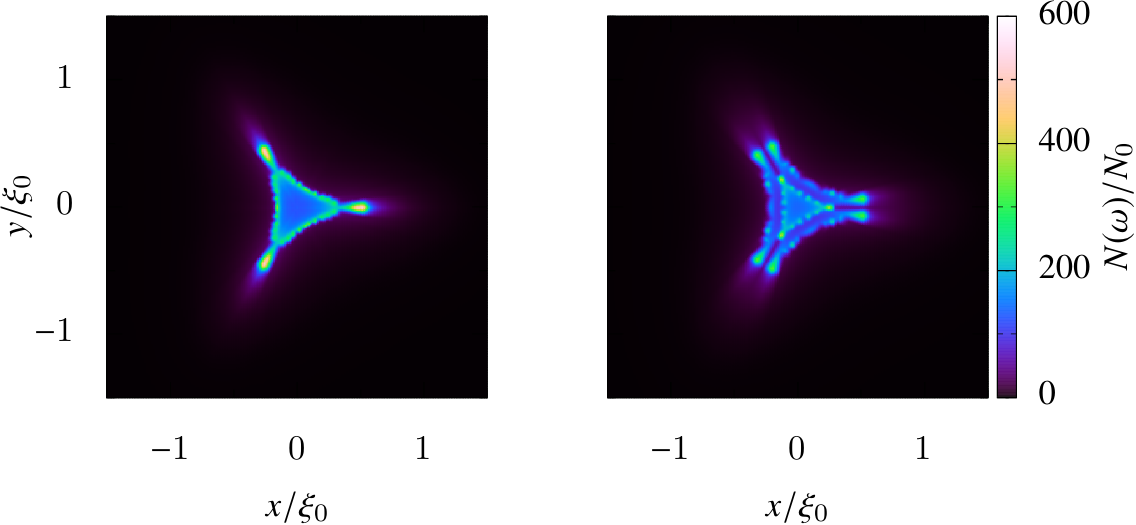}
\caption{(Color online) The LDOS $N(\omega,\bm{r})=-\Im \average{g(\omega+\ki\eta,\hat k,\bm{r})}_{\hat k}/\kpi$ around the non-axisymmetric triangular parallel vortex in 2D chiral $p$-wave superconductors at \(T/\tc=0.2\). We set the smearing factor $\eta=0.01\kk\tc/\hbar$ and use 4800 equally spaced points of momentum to integrate over the Fermi surface. Left: $\hbar\omega = 0.0\kk\tc$, right: $\hbar\omega = 0.3\kk\tc$.}
\label{fig:non-axisymmetric-vortex-ldos}
\end{figure}

\begin{figure}
\centering
\includegraphics[width=.96\columnwidth]{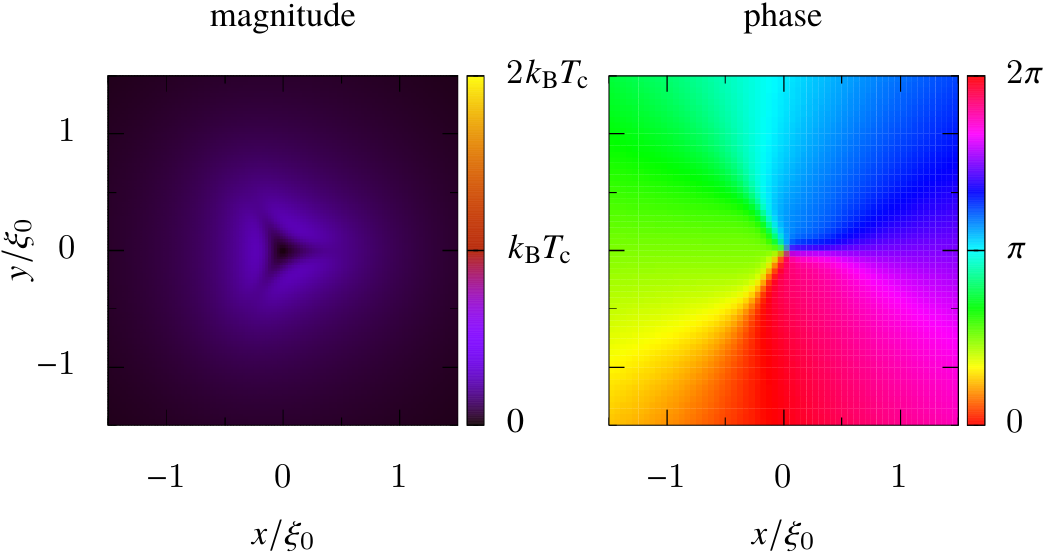}
\caption{(Color online) Self-energy (\(\Sigma_{+}\)) around the non-axisymmetric triangular parallel vortex in 2D chiral $p$-wave superconductors at \(T/\tc=0.2\).}
\label{fig:non-axisymmetric-vortex-fock}
\end{figure}

In Figs.~\ref{fig:non-axisymmetric-vortex-major} and \ref{fig:non-axisymmetric-vortex-minor}, we respectively show the solutions of the major and minor components of the pair potentials of a parallel vortex in chiral $p$-wave superconductors. As can be seen, the vortex breaks the axisymmetry, and the vortex of the minor component splits into three.
We cannot obtain this highly non-axisymmetric solution when we exclude the self-energy term \(\check\Sigma\) from the gap equations. Therefore, we conclude that the emergence of the non-axisymmetric (meta)stable vortex reflects the existence of the self-energy. 
We plot the magnetic field and the electric current density in Fig.~\ref{fig:nonaxisymmetric-electromagnetic}. We can see that the axisymmetry is broken and a triangular shape is formed in both quantities. We plot the local density of states (LDOS) around the vortex in Fig.~\ref{fig:non-axisymmetric-vortex-ldos} and the self-energy in Fig.~\ref{fig:non-axisymmetric-vortex-fock}; they also have triangular profiles (see also Appendix~\ref{appendix:local-density-of-states}).

The solution depends on the initial condition of the self-consistent calculation; the solution is non-axisymmetric if we set the initial pair potentials to be sufficiently non-axisymmetric, and becomes axisymmetric otherwise, at $T/\tc\le 0.45$. The non-axisymmetric results for various initial conditions have the same shape for a given temperature; this implies that the non-axisymmetric vortices are at least metastable. To compare the relative stabilities of the axisymmetric and non-axisymmetric solutions, we calculate the free energies of the vortices.

\begin{figure}
\centering
\includegraphics[width=.8\columnwidth]{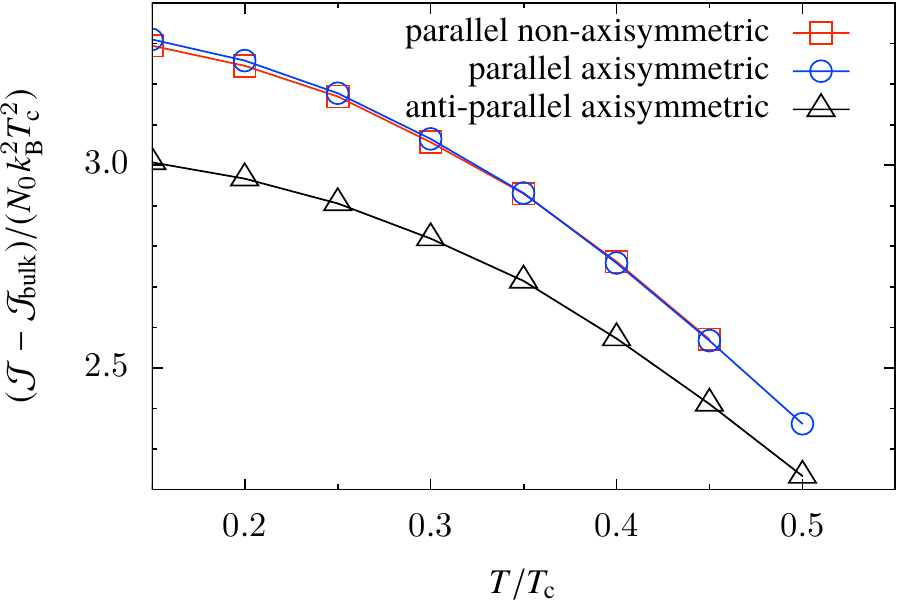}
\caption{(Color online) Free energy \(\mathcal{J}\) of an isolated vortex in 2D chiral $p$-wave superconductors. The reference level of the free energy is its bulk \(\mathcal{J}_{\text{bulk}}\). red squares: parallel non-axisymmetric; blue circles: parallel axisymmetric; black triangles:  antiparallel vortex. }
\label{fig:free-energy-vs-temperature}
\end{figure}

\begin{figure}
\centering
\includegraphics[width=.84\columnwidth]{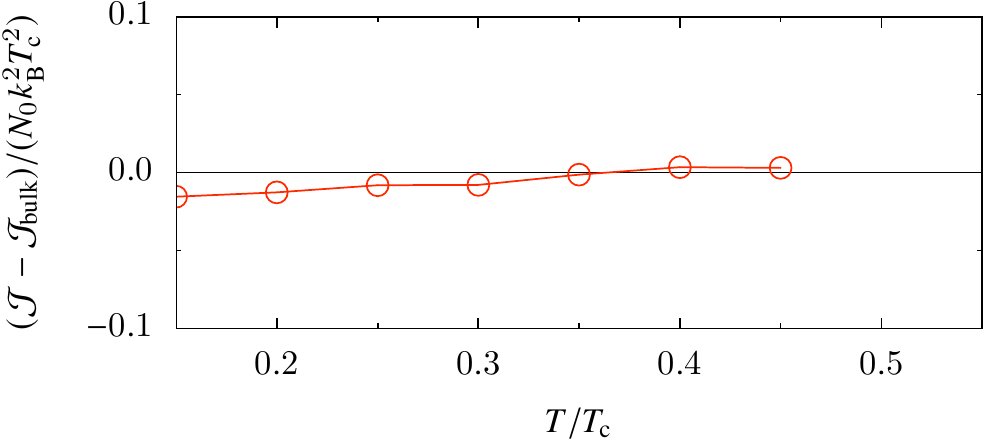}
\caption{(Color online) Difference in free energies between axisymmetric and non-axisymmetric parallel vortices in 2D chiral $p$-wave superconductors (\(\upDelta\mathcal{J}=\mathcal{J}_{\text{non-axisymmetric}} - \mathcal{J}_{\text{axisymmetric}}\)).}
\label{fig:free-energy-diff}
\end{figure}

Figure~\ref{fig:free-energy-vs-temperature} shows the free energy of each vortex and Fig.~\ref{fig:free-energy-diff} shows the difference in free energies between axisymmetric and non-axisymmetric vortices. We do not find the non-axisymmetric vortex as a metastable state at \(T/\tc=0.5\). We can see that both axisymmetric and non-axisymmetric parallel vortices have larger free energies than the (axisymmetric) antiparallel vortices; the parallel vortex is still metastable. Among the metastable states, the non-axisymmetric vortices are stabler than, or at least as stable as, the axisymmetric vortices at lower temperatures (the crossover temperature is about \(0.35\tc\)).

The importance of the non-axisymmetric vortex is not limited to theoretical aspects.
The ground states of the chiral superconductors break the time-reversal symmetry and are degenerated without external magnetic fields. The chiral $p$-wave superconductors, therefore, are believed to form structures of domains corresponding to each degenerated state.
At the moment, there are two systems in which chiral $p$-wave superconductivity/superfluidity is known to be realized: A-phase of superfluid helium-3\cite{Wheatley1975,Legett1975} and perhaps \ce{Sr2RuO4}\cite{Maeno1994,Mackenzie2003,Sigrist2005,Maeno2012}.
While chiral domain structures have been observed in the \ce{^3He} A-phase\cite{Ikegami2013,Ikegami2015}, there has been no report of direct observation of chiral domains in \ce{Sr2RuO4}. 
The results presented in Sect.~4 imply that we can obtain evidence of chiral domains by observing both axisymmetric vortices and threefold rotational symmetric vortices in a sample of chiral $p$-wave superconductor at sufficiently low temperature and with external magnetic fields\cite{footnote-domains}, e.g., through LDOS measurements by scanning tunneling microscopy (STM).

Tokuyasu \textit{et al.} studied the vortex in chiral $p$-wave superconductivity on the basis of the Ginzburg--Landau theory and reported that the vortex spontaneously breaks the axisymmetry within some parameter region\cite{Tokuyasu1990}. Although their profiles of the pair potentials appear similar to ours, our obtained vortex is, however, essentially different from theirs for the following reasons.

According to the Ginzburg--Landau theory in a previous study\cite{Tokuyasu1990}, the vortex in the system at the weak-coupling limit does not break the axisymmetry. The presence of the self-energy term makes our system different from the conventional weak-coupling systems. This difference, however, disappears near the limit of \(T\!\nearrow\!\tc\) and does not affect the Ginzburg--Landau equation (see Appendix~\ref{appendix:fock-term-and-ginzburg-landau-equation}); the present system does not break the axisymmetry in the Ginzburg--Landau theory. This is also consistent with the fact that the non-axisymmetric vortex becomes unstable at \(T/\tc \gtrsim 0.35\) in the present study. 
In addition, with the setup in this study, we do not find any non-axisymmetric antiparallel vortices, which where  obtained when using the Ginzburg--Landau theory. With this reasoning, we consider that the breaking of the symmetry in this work cannot be explained by the Ginzburg--Landau theory and is an unprecedented phenomenon.

One of the reasons for the breaking of the axisymmetry is the existence of multi-winding induced components of the order parameters. In general, the energy of a multi-winding vortex is higher than the sum of the energies of single-winding vortices. In some systems, the axisymmetry of vortices is broken in the manner of the splitting of multi-winding vortices of induced components\cite{Fogelstroem1995,Silaev2015,Tokuyasu1990,Ogawa2000,Thuneberg1986,Salomaa1986,Fogelstroem1999}. In these cases, however, the vortex splitting depends on the parameters of the system; in one parameter regime, the axisymmetry is broken, and in another regime, it is not. Thus, the exact conditions and physics of these breakings of axisymmetry are not yet fully revealed.

%Another important point of our result is the effect of the diagonal self-energy. We point out that a similar (but not the entirely same) effect has been reported on the B-phase of superfluid helium-3. In this system, there are two types of stable vortices\cite{Thuneberg1986,Salomaa1986,Salomaa1983}: the axisymmetric  and the non-axisymmetric \textit{$v$-vortices}. The quasiclassical theory has predicted that the existence of the diagonal self-energy changes the transition temperature between the vortices\cite{Fogelstroem1995,Silaev2015,footnote-helium3-b-phase-vortices}. Thus, it is not hard to realize that the self-energy changes the condition of transition between vortices. The detailed mechanisms of the changes in both systems are issues in the future.

Another important point of our result is the effect of the diagonal self-energy. We point out that a similar (but not entirely the same) effect has been reported for the B-phase of superfluid helium-3. In this system, there are two types of stable vortices\cite{Thuneberg1986,Salomaa1986,Salomaa1983}: the axisymmetric and non-axisymmetric \textit{$v$-vortices}. There also exists most symmetric but metastable \textit{o-vortex}. A study using the quasiclassical theory\cite{Fogelstroem1995} reported that when changing the diagonal self-energy, though the relative stability does not change, changes of free energies of the three vortices are different from each other. Thus, it is not difficult to realize that the self-energy changes the condition of the transition between vortices. The detailed mechanisms of the changes in these systems are issues in the future.

%%%%%%%%%%%%%%%%%%%%%%%%%%%%%%%%%%%%%%%%%%%%%%%%%%%%%%%%%%%%%%%%%%%%%%%
\section{Conclusion}

In the present paper, we consider the weak-coupling chiral superconductors whose frequency dependence of order parameters is ignorable. We study an isolated vortex with the diagonal self-energy corresponding to the Fock term, which has been paid little attention so far. We confirm that the term does not come out within vortices of even-parity superconductors. On the other hand, we find that a finite-magnitude self-energy emerges within vortices of odd-parity superconductors, particularly \(p\)-wave superconductors in general, even at the quasiclassical level.
It is considered not to be rare that the self-energy term is renormalizable and has no significant effect on the vortex. However, at least within \textit{parallel vortices} in the chiral $p$-wave superconductor, this term causes the broken axisymmetry of the vortex, which cannot be described by the Ginzburg--Landau theory.
The above implies that the self-energy term is not negligible within vortices in odd-parity superconductors in general, even in the weak-coupling limit.

In this paper, we only considered the systems with one spin component, e.g., \(\bm{d}= (\hat k_x\pm\ki \hat k_y)\hat z\), or systems in which the Eilenberger equation and the gap equation are decoupled into two or more spin components and we can solve each spin component independently, e.g., \(\bm{d}= \hat k_x \hat x + \hat k_y \hat y\). Of course, we cannot always decouple spin components. For example, the d-vector in the superfluid helium-3 B-phase has the form \(\bm{d}=\hat k_x\hat x+\hat k_y\hat y+ \hat k_z\hat z\). Extension to systems with multicomponent spin is a problem for future study.

These days, many noncentrosymmetric superconductors have been discovered and have attracted much attention\cite{Smidman2017}. Within these systems, the parity is not a good quantum number, and in superconducting states of such a system, singlet Cooper pairs and triplet pairs are mixed.
Within a vortex in such systems, the self-energy term also cannot be negligible in general, as long as a sufficiently large \(p\)-wave component exists. 
It is important to study the effect of the self-energy term (with or without frequency dependences) on these systems for the purpose of further understanding the physics of odd-parity superconductivity.

%%%%%%%%%%%%%%%%%%%%%%%%%%%%%%%%%%%%%%%%%%%%%%%%%%%%%%%%%%%%%%%%%%%%%%%
%\begin{acknowledgments}
\section*{Acknowledgments}
We would like to thank Y. Kato, Y. Tsutsumi, M. Ichioka, J. A. Sauls, G. E. Volovik, and E. V. Thuneberg for their helpful discussions and encouragement. This work was supported by JSPS KAKENHI Grant Number 15K05160.
%\end{acknowledgments}

%%%%%%%%%%%%%%%%%%%%%%%%%%%%%%%%%%%%%%%%%%%%%%%%%%%%%%%%%%%%%%%%%%%%%%%
\appendix

%%%%%%%%%%%%%%%%%%%%%%%%%%%%%%%%%%%%%%%%%%%%%%%%%%%%%%%%%%%%%%%%%%%%%%%
\section{Details of Numerical Calculation}\label{appendix:numerical}
In Sect.~3 and 4, we carried out the numerical calculation of an isolated vortex in chiral superconductors. In this appendix, we describe the details of the calculation. We use a 2D polar coordinate system, and choose the center of the vortex  as the origin.
The sampling point on the spatial coordinates and the azimuthal angle of the momentum $\phi$ are the same as in our previous proceeding \cite{Kurosawa2017}. For 3D chiral superconductors, we take 12 points equally spaced on the polar angle of the momentum $\theta$.

The specific forms of coupling function in the gap equation \(v(\hat k,\hat k')\) are given as
\begin{align}
& 2 c_{0}\cos(\phi-\phi') &\text{(chiral $p$)}
,\\
& 2 c_{0}\cos[2(\phi-\phi')] &\text{(chiral $d$)}
,\\
& 2 c_{0}\cos[3(\phi-\phi')] &\text{(chiral $f$)}
\end{align}
for 2D chiral superconductors,
\begin{align}
& 4\kpi c_{0}\sum_{m=\pm 1}Y_{1,m}(\theta,\phi)Y_{1,m}^{*}(\theta',\phi') &\text{(chiral $p$)}
,\\
& 4\kpi c_{0}\sum_{m=\pm 1}Y_{2,m}(\theta,\phi)Y_{2,m}^{*}(\theta',\phi') &\text{(chiral $d$ ($m=1$))}
,\\
& 4\kpi c_{0}\sum_{m=\pm 1}Y_{3,m}(\theta,\phi)Y_{3,m}^{*}(\theta',\phi') &\text{(chiral $f$ ($m=1$))}
,\\
& 4\kpi c_{0}\sum_{m=\pm 3}Y_{3,m}(\theta,\phi)Y_{3,m}^{*}(\theta',\phi') &\text{(chiral $f$ ($m=3$))}
,
\end{align}
for 3D systems, and $v(\hat k,\hat k')=c_0$ for $s$-wave superconductors, where
\begin{align}
\frac{1}{N_0 c_{0}} &= 2\kpi\kk T\sum_{0 <\epsilon_n\le\epsilon_{\text{c}}}\frac{1}{\hbar\epsilon_n} + \ln\frac{T}{T_{\text{c}}}
,
\end{align}
and we take \(\epsilon_{\text{c}}=20 \kk \tc/\hbar\).

We can reduce the Eilenberger equation [Eq.~\eqref{eq:eilenberger-equation}] to two independent decoupled Riccati-type ordinary differential equations\cite{Nagato1993,Schopohl1995,Eschrig1999},
\begin{subequations}
\begin{align}
-\bm{v}_{\text{F}}\cdot(-\ki\hbar\nabla-2q\bm{A})\gamma &= -\Delta^{*}\gamma^2 - 2(\hbar z-\Sigma)\gamma - \Delta
,
\\
-\bm{v}_{\text{F}}\cdot(-\ki\hbar\nabla+2q\bm{A})\bar\gamma &= -\Delta\bar\gamma^2 + 2(\hbar z-\Sigma)\gamma - \Delta^{*}
,
\end{align}
\label{eq:riccaci-parametrized-equation}%
\end{subequations}
and the quasiclassical Green's function \(\check g\) is expressed in terms of \(\gamma\) and \(\bar\gamma\) as
\begin{align}
\check g &= \frac{-\ki\kpi\sgn\epsilon_n}{1+\gamma\bar\gamma}\begin{pmatrix} 1-\bar\gamma\gamma & 2\gamma \\ 2\bar\gamma & -1+\bar\gamma\gamma \end{pmatrix}
.
\end{align}
This parametrization is often called the \textit{Riccati parametrization}. This parametrization gives us a stable numerical calculation\cite{Nagai2012} and a clear view into the perturbative analysis of the Green's function.
We use a fourth- and fifth-order adaptive Runge-Kutta method\cite{Shampine1986} to solve Eq.~\eqref{eq:riccaci-parametrized-equation}. 

We solve the Eilenberger equation, the gap equation, and the Maxwell--Ampère equation with the Coulomb (London) gauge self-consistently. We repeat the self-consistent loop until the maximum difference between the new and old \(\check\Sigma\) is smaller than \(5\times10^{-5}\kk\tc\). To improve the speed and stability of the convergence, we use a variant of the Anderson acceleration method\cite{Eyert1996}.

%%%%%%%%%%%%%%%%%%%%%%%%%%%%%%%%%%%%%%%%%%%%%%%%%%%%%%%%%%%%%%%%%%%%%%%
\section{Quasiclassical Green's Function}\label{appendix:quasiclassical}
The quasiclassical Green's function\cite{Eilenberger1968,Larkin1969} is defined as follows.
First, we define the equilibrium Nambu--Gor'kov Green's function. We use $\vec x$ as an abbreviation for $(\bm{x},\tau,\sigma)$, where $\bm{x}$ denotes the space coordinate,  $\tau$ denotes the imaginary time, and $\sigma$ denotes the spin. The Green's function $\check G(\vec{x},\vec{x}')$ can be written as
\begin{align}
-\check G(\vec{x},\vec{x}') = \begin{pmatrix} 
\daverage{\operatorname{T}\mfc(\vec{x})\hconj{\mfc}(\vec{x}')} & \daverage{\operatorname{T}\mfc(\vec{x})\mfc(\vec{x}')}
\\
\daverage{\operatorname{T}\hconj{\mfc}(\vec{x})\hconj{\mfc}(\vec{x}')} & \daverage{\operatorname{T}\hconj{\mfc}(\vec{x})\mfc(\vec{x}')}
\end{pmatrix}
,
\end{align}
where $\operatorname{T}$ is the time-ordered product, and $\hconj{\mfc}(\vec{x})$ and $\mfc(\vec{x})$ are the creation and annihilation operators of the fermions, respectively. We define $\daverage{X}=\tr\{\rho X\}$, where $\rho$ is the density matrix in equilibrium.

Second, we write $\check G(\vec{x},\vec{x}')$ as a function of the center of the space coordinate $\bm{r}$, wave vectors of the relative space coordinate $\bm{k}$, and the Matsubara frequencies $\ki\epsilon_n$. With the Wigner transformation, we define
\begin{align}
\check G(\vec{x},\vec{x}') = \frac{\kk T}{\hbar}\sum_n\int\frac{\dd\bm{k}}{(2\kpi)^3}\check G(\ki\epsilon_n,\bm{k},\bm{r})\ke^{-\ki\epsilon_n\bar\tau+\ki\bar{\bm{r}}\bm{k}}
,
\end{align}
where $\bm{r} = (\bm{x}+\bm{x}')/2$, $\bar{\bm{r}} = \bm{x}-\bm{x}'$, and $\bar\tau = \tau - \tau'$.

Finally, we integrate $\check G$ over the energy $\xi_{\bm{k}}$ to concentrate on the properties just upon the Fermi level. In the present paper, we assume that the system has a spherical (circular) Fermi surface; in this case, we can take $\xi_{\bm{k}} = \hbar^2k^2/(2m^{*}) - \mu_{\text{F}}$, where $m^{*}$ is the effective mass of the particle and $\mu_{\text{F}}$ is the Fermi energy. We define the quasiclassical Green's function $\check g$ as
\begin{align}
\hbar\check g(\ki\epsilon_n,\hat k,\bm{r}) &= \int\dd\xi_{\bm{k}} \breve\tau_3\check G(\ki\epsilon_n,\bm{k}, \bm{r})
,
\end{align}
where $\hat k=\bm{k}/\abs{k}$ and
\begin{align}
\breve\tau_3 = \begin{pmatrix} \check\tau_0 & 0 \\ 0 & -\check\tau_0 \end{pmatrix}.
\end{align}
We also define $g$, $\bar g$, $f$, $\bar f$ as
\begin{align}
\begin{pmatrix} g(\ki\epsilon_n,\hat k,\bm{r}) & f(\ki\epsilon_n,\hat k,\bm{r}) \\
 -\bar f(\ki\epsilon_n,\hat k,\bm{r}) & \bar g(\ki\epsilon_n,\hat k,\bm{r}) \end{pmatrix} = \check g(\ki\epsilon_n,\hat k,\bm{r})
 \label{eq:2x2-2x2 quasiclassical Green's function definition}
.
\end{align}
In the above equation, $g(\ki\epsilon_n,\hat k,\bm{r})$ is a $2\times 2$ matrix in spin space, and so are $\bar g$, $f$ and $\bar f$.
If the particle-hole symmetry holds, we have $\bar g$ = $-g$ as well (see Ref.~\citen{Kopnin2001}, for example).
Taking the analytic continuation from imaginary-frequencies $\ki\epsilon_n$ with $\epsilon_n>0$ to real frequencies $\omega+\ki\eta$, we can obtain the quasiclassical retarded Green's function, where $\eta$ is a positive infinitesimal.

%%%%%%%%%%%%%%%%%%%%%%%%%%%%%%%%%%%%%%%%%%%%%%%%%%%%%%%%%%%%%%%%%%%%%%%
\section{Low-Energy States in Non-Axisymmetric Vortex}\label{appendix:local-density-of-states}

In Sect.~4, we find an exotic pattern of zero-energy LDOS. In this appendix, we try to explain this pattern in terms of the well-known Andreev bound states (ABS) within a vortex\cite{Andreev1964,Kulik1970,Stone1996}.

One of the simplest views of the vortex core is to consider it as a superconductor-normal-superconductor (SNS) junction\cite{Stone1996}. In the SNS junction, there exist low-energy states (i.e., ABS)\cite{Kulik1970}, and the phase difference between the two sides of the superconductor is crucially important for these states. If the phase difference is \(\kpi\), the energy of the ABS becomes zero. In the case of the vortex as a normal core, the path where the phase difference between the two superconducting edges is \(\kpi\) is the path running across the exact center of the vortex. This view is successful for the vortex in various systems, as well as conventional superconductors\cite{Stone1996}.

We now return to our system. At first glance, the quasiclassical path of the quasiparticles of zero energy does not  obey the above rule and the simple view of the vortex seems to fail. However, once we assume the spatial profiles of the pair potentials, we can show that the above view still holds in some aspects.

\begin{figure*}
\centering
\begin{minipage}{.45\textwidth}\centering
\includegraphics[width=\textwidth]{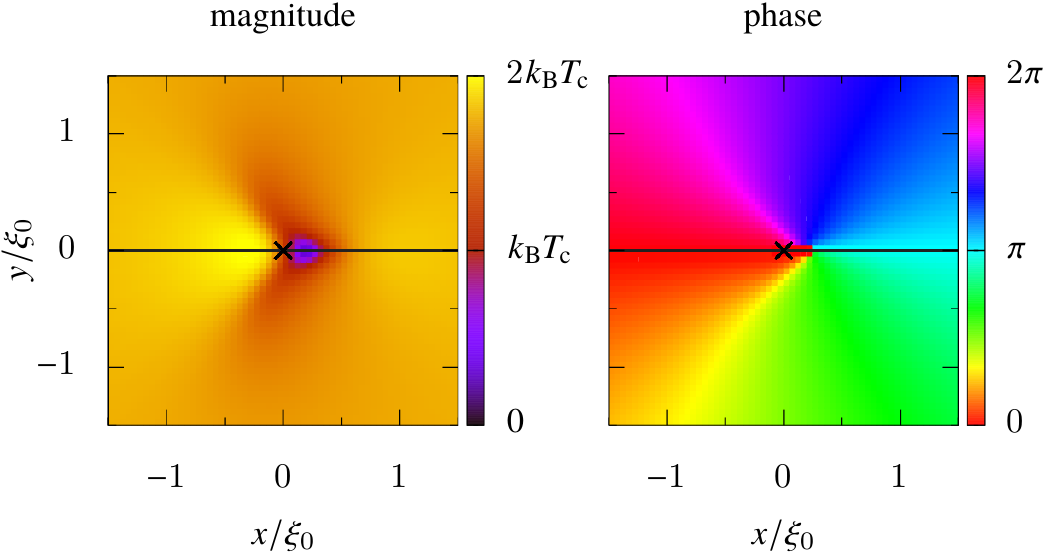}
{(a): \(\phi=0\)}
\end{minipage}
\begin{minipage}{.45\textwidth}\centering
\includegraphics[width=\textwidth]{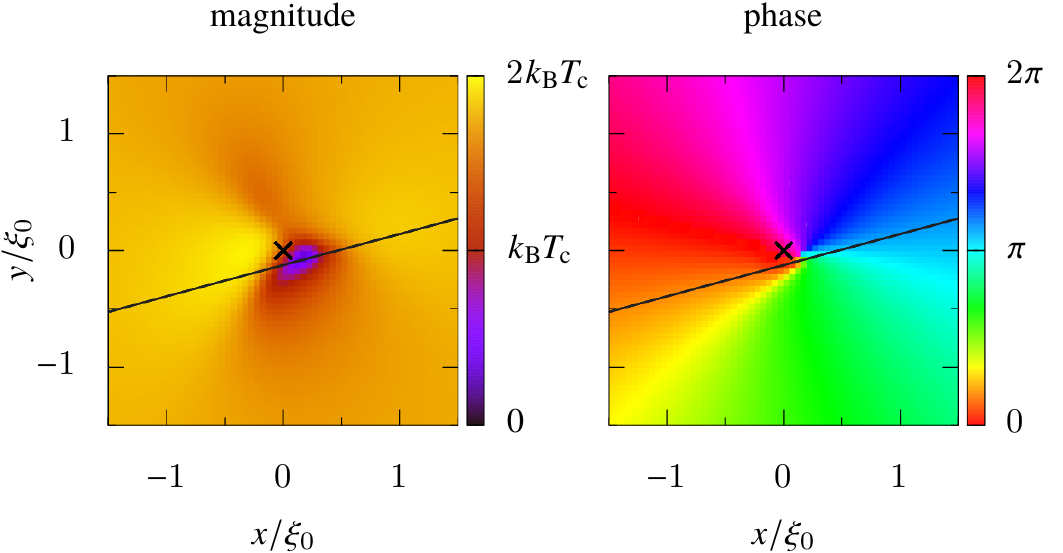}
{(b): \(\phi=\kpi/12\)}
\end{minipage}

\vspace{2ex}
\begin{minipage}{.45\textwidth}\centering
\includegraphics[width=\textwidth]{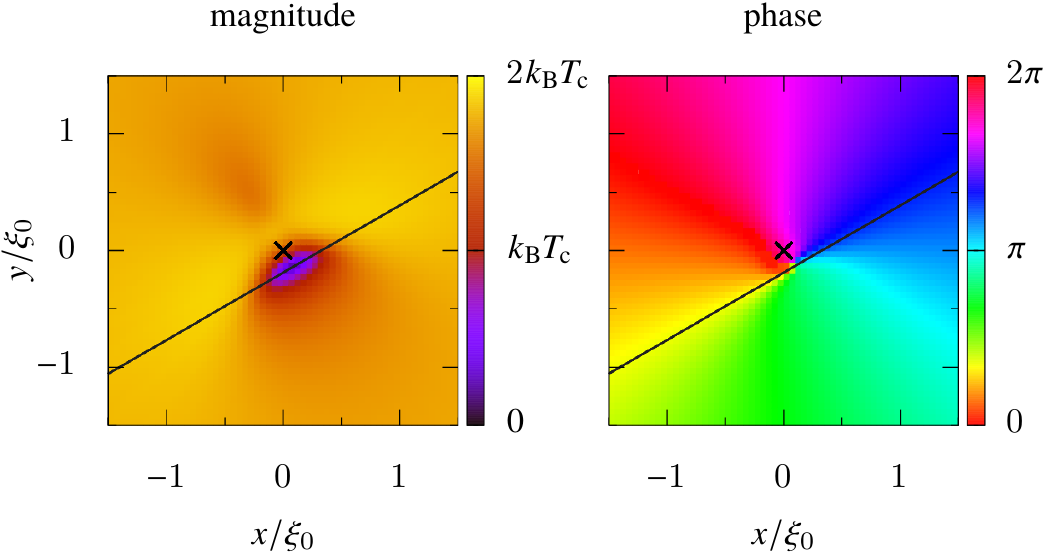}
{(c): \(\phi=\kpi/6\)}
\end{minipage}
\begin{minipage}{.45\textwidth}\centering
\includegraphics[width=\textwidth]{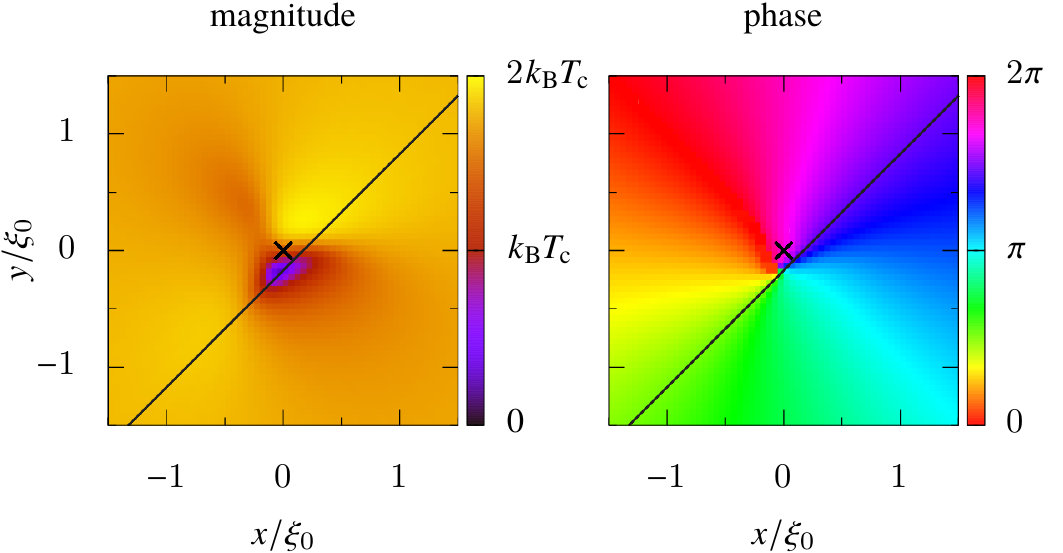}
{(d): \(\phi=\kpi/4\)}
\end{minipage}

\caption{
  (Color online)
  Pair potentials of a non-axisymmetric parallel vortex for some momenta (\(T/\tc=0.2\)). The solid line is the trajectory of the quasiparticle forming a zero-energy bound state obtained by the Eilenberger equations. We also placed a cross at the center of the system (the center of the vortex of the major component).}
\label{fig:pair-potential-for-each-momentum}
\end{figure*}

We emphasize that the chiral $p$-wave system contains two components of pair potentials (because it is ``chiral'') and each part has a different dependence on the momentum (because of ``$p$-wave''). For each quasiparticle, the pair potential in its equations of motion is neither of the isolated component but the sum of the components. Figure~\ref{fig:pair-potential-for-each-momentum} shows the pair potentials for some momenta.
We see that the \textit{center} of the vortex for each quasiparticle is not the same as the center of the system, owing to the minor components of the pair potential. As we expected, the path with zero energy ABS runs near the \textit{effective vortex center}, where the phases of the pair potentials at both sides of a normal region change \(\kpi\).
Once we obtain the paths of ABS, we can reproduce the LDOS pattern as its enveloping path, as successfully carried out in various systems\cite{Stone1996,Ichioka1996,Nagai2006}.

%% one-column

%%%%%%%%%%%%%%%%%%%%%%%%%%%%%%%%%%%%%%%%%%%%%%%%%%%%%%%%%%%%%%%%%%%%%%%
\section{Diagonal Self-Energy and Ginzburg--Landau Equations}
\label{appendix:fock-term-and-ginzburg-landau-equation}

There is a standard method of deriving the Ginzburg--Landau equations from the quasiclassical Eilenberger equations and the gap equations\cite{Kopnin2001}. In this methodology, we solve the quasiclassical Green's function near the critical temperature by the perturbation of \(\Delta/(\kk T) \sim \Delta/(\hbar\epsilon_n)\), substitute the solutions into the gap equations, and derive the equations of the pair potentials. 

The perturbative solution of quasiclassical Green's functions near the critical temperature is
\begin{align}
g &=
-\ki\pi\sgn\epsilon_n
+\frac{\ki\pi\abs{\Delta}^2}{2\hbar\tilde\epsilon_n\abs{\hbar\tilde\epsilon_n}}
\nonumber\\
&\qquad+
\frac{\ki\pi\hbar}{4\abs{\hbar\tilde\epsilon_n}^3}[
  -\Delta(\bm{v}_{\text{F}}\cdot\nabla_{+})\Delta^{*}
  +\Delta^{*}(\bm{v}_{\text{F}}\cdot\nabla_{-})\Delta
  ]
\label{eq:GL-perturbative-solution-g}
,\\
f &=
\frac{\pi\Delta}{\abs{\hbar\tilde\epsilon_n}}
-\frac{\pi\hbar(\bm{v}_{\text{F}}\cdot\nabla_{-})\Delta}{2\hbar\tilde\epsilon_n\abs{\hbar\tilde\epsilon_n}}
-\frac{\pi\Delta\abs{\Delta}^2}{2\abs{\hbar\tilde\epsilon_n}^3}
+\frac{\pi\hbar^2(\bm{v}_{\text{F}}\cdot\nabla_{-})^2\Delta}{4\abs{\hbar\tilde\epsilon_n}^3}
,\\
\bar f &=
\frac{\pi\Delta^{*}}{\abs{\hbar\tilde\epsilon_n}}
+\frac{\pi\hbar(\bm{v}_{\text{F}}\cdot\nabla_{+})\Delta^{*}}{2\hbar\tilde\epsilon_n\abs{\hbar\tilde\epsilon_n}}
-\frac{\pi\Delta^{*}\abs{\Delta}^2}{2\abs{\hbar\tilde\epsilon_n}^3}
+\frac{\pi\hbar^2(\bm{v}_{\text{F}}\cdot\nabla_{+})^2\Delta^{*}}{4\abs{\hbar\tilde\epsilon_n}^3}
,
\end{align}
where \(\hbar\tilde\epsilon_n = \hbar\epsilon_n + \ki\Sigma\) and \(\nabla_{\mp} = \nabla\mp2\ki q\bm{A}/\hbar\).
Because the Ginzburg--Landau equations are third-order equations of the order parameters, we only consider the zeroth to the third-order terms in the solution of \(\check g\).

Substituting Eq.~\eqref{eq:GL-perturbative-solution-g} into Eq.~\eqref{eq:weak-coupling-eliashberg-equation-diagonal}, we can see that \(\Sigma\) is \(\mathrm{O}\left([\abs{\Delta}/(\hbar\epsilon_n)]^3\right)\). This means that the effect of \(\Sigma\) on \(f\) is of the fourth order of the perturbation. Thus, the diagonal part of the self-energy does not appear in the Ginzburg--Landau equation of the pair potentials. Also, the leading term related to \(\Sigma\) is on the order of \(\Sigma^2\) because there is no self-energy term in the third order equations for pair potentials. In addition, the magnitude of \(\Sigma^2\) is of sixth order of the perturbation; terms such as \(\Sigma f\) and \(\Sigma^2\) are not in the Ginzburg--Landau free energy functional. Therefore, the diagonal self-energy does not change the Ginzburg--Landau equation up to the third order term of the pair potential.


\begin{thebibliography}{99}
\bibitem{Migdal1958} A.\ B.\ Migdal, Sov.\ Phys. JETP \textbf{7}, 996 (1958). 
\bibitem{Eliashberg1960} G.\ M.\ Eliashberg, Sov.\ Phys.\ JETP \textbf{11}, 696 (1960).
\bibitem{Morel1962} P.\ Morel and A.\ W.\ Anderson, Phys.\ Rev.\ \textbf{125}, 1263 (1962).
\bibitem{Bardeen1957} J.\ Bardeen, L.\ N.\ Cooper, and J.\ R.\ Schrieffer, Phys.\ Rev.\ \textbf{108}, 1175 (1957).
\bibitem{Scalapino1969} \textit{Superconductivity}, ed.\ R.\ D.\ Parks (Marcel Dekker, New York, 1969) Chap.\ 10.
\bibitem{Caroli1964} C.\ Caroli, P.\ G.\ de~Gennes, and J.\ Matricon, Phys.\ Lett.\ \textbf{9}, 307 (1964).
\bibitem{Kopnin1976} N.\ B.\ Kopnin and V.\ E.\ Kravtsov, JETP Lett.\ \textbf{23}, 578 (1976).
\bibitem{Kopnin1991} N.\ B.\ Kopnin and  M.\ M.\ Salomaa, Phys.\ Rev.\ B \textbf{44}, 9667 (1991).
\bibitem{Kopnin1997} N.\ B.\ Kopnin and G.\ E.\ Volovik, Phys.\ Rev.\ Lett.\ \textbf{79}, 1377 (1997).
\bibitem{Fogelstroem1995} M.\ Fogelström and J.\ Kurkijärvi, J.\ Low Temp.\ Phys.\ \textbf{98}, 195 (1995); \textbf{100}, 597(E) (1995). 
\bibitem{Silaev2015} M.\ A.\ Silaev, E.\ V. Thuneberg, and M.\ Fogelström, Phys.\ Rev.\ Lett.\ \textbf{115}, 235301 (2015).
\bibitem{McMillan1968} W.\ L.\ McMillan, Phys.\ Rev.\ \textbf{175}, 537 (1968).
\bibitem{Kurosawa2017} N.\ Kurosawa and Y.\ Kato, J.\ Low Temp.\ Phys.\ \textbf{187}, 538 (2017).
\bibitem{Eilenberger1968} G.\ E.\ Eilenberger, Z.\ Phys.\ \textbf{214}, 195 (1968).
\bibitem{Larkin1969} A.\ I.\ Larkin and Yu. N. Ovchinnikov, Sov.\ Phys.\ JETP \textbf{28}, 1200 (1969).
\bibitem{Eliashberg1963} G.\ M.\ Eliashberg, Sov.\ Phys.\ JETP \textbf{16}, 780 (1963).
\bibitem{Bardeen1964} J.\ Bardeen and M.\ Stephen, Phys.\ Rev.\ \textbf{136}, A1485 (1964).
\bibitem{Serene1983} J.\ W.\ Serene and D.\ Rainer, Phys.\ Rep.\ \textbf{101}, 221 (1983).
\bibitem{Thuneberg1984} E.\ V.\ Thuneberg, J.\ Kurkijärvi, and D.\ Rainer, Phys. Rev. B \textbf{29}, 3913 (1984).
\bibitem{Kramer1974} L.\ Kramer and W.\ Pesch, Z.\ Phys.\ \textbf{269}, 59 (1974).
\bibitem{Kato2000} Y.\ Kato, J.\ Phys.\ Soc.\ Jpn.\ \textbf{69}, 3378 (2000).
\bibitem{Nagai2006} Y.\ Nagai, Y.\ Ueno, Y.\ Kato, and N.\ Hayashi, J.\ Phys.\ Soc.\ Jpn.\ \textbf{75}, 104701 (2006).
\bibitem{Matsumoto1999} M.\ Matsumoto and M.\ Sigrist, J.\ Phys.\ Soc.\ Jpn.\ \textbf{68}, 724 (1999).
\bibitem{Hayashi2005} N.\ Hayashi, Y.\ Kato, and M.\ Sigrist, J.\ Low Temp.\ Phys.\ \textbf{139}, 79 (2005).
\bibitem{Tanuma2009} Y.\ Tanuma, N.\ Hayashi, Y.\ Tanaka, and A.\ A.\ Golubov, Phys.\ Rev.\ Lett.\ \textbf{102}, 117003 (2009).
\bibitem{Sauls2009} J.\ A.\ Sauls and M.\ Eschrig, New J.\ Phys.\ \textbf{11}, 075008 (2009).
\bibitem{Wheatley1975} J.\ C.\ Wheatley, Rev.\ Mod.\ Phys.\ \textbf{47}, 415 (1975).
\bibitem{Legett1975} A.\ J.\ Leggett, Rev.\ Mod.\ Phys.\ \textbf{47}, 331 (1975).
\bibitem{Maeno1994} Y.\ Maeno, H.\ Hashimoto, K.\ Yoshida, S.\ Nishizaki, T.\ Fujita, J.\ G.\ Bednorz, and F.\ 
Lichtenberg, Nature \textbf{372}, 532 (1994).
\bibitem{Mackenzie2003} A.\ P.\ Mackenzie and Y.\ Maeno, Rev.\ Mod.\ Phys.\ \textbf{75}, 657 (2003).
\bibitem{Sigrist2005} M.\ Sigrist, Prog.\ Theor.\ Phys.\ Suppl.\ \textbf{160}, 1 (2005).
\bibitem{Maeno2012} Y.\ Maeno, S.\ Kittaka, T.\ Nomura, S.\ Yonezawa, and K.\ Ishida, J.\ Phys.\ Soc.\ Jpn.\ \textbf{81}, 011009 (2012).
\bibitem{Ikegami2013} H.\ Ikegami, Y.\ Tsutsumi, and K.\ Kono, Science \textbf{341}, 59 (2013).
\bibitem{Ikegami2015} H.\ Ikegami, Y.\ Tsutsumi, and K.\ Kono, J.\ Phys.\ Soc.\ Jpn.\ \textbf{84}, 044602 (2015).

\bibitem{footnote-domains} {Since the antiparallel vortices are stabler than the parallel vortices, the domain structures are bound to disappear under magnetic fields. In real systems, however, there are always defects that pin the domain walls; when there are few perturbations that can overcome the pinning, domains may survive for certain hours under even the existence of external magnetic fields. Thus, we may obtain parallel vortices as follows: first, cool down the system without fields to obtain the superconductor with many domains, and second, add fields.}


\bibitem{Tokuyasu1990} T.\ A.\ Tokuyasu, D.\ W.\ Hess, and J.\ A.\ Sauls, Phys.\ Rev.\ B \textbf{41}, 8891 (1990).
\bibitem{Ogawa2000} N.\ Ogawa and M.\ E.\ Zhitomirsky, J.\ Phys.\ Soc.\ Jpn.\ \textbf{69}, 3660 (2000).

\bibitem{Thuneberg1986} E.\ V.\ Thuneberg, Phys.\ Rev.\ Lett.\ \textbf{56}, 359 (1986).
\bibitem{Salomaa1986} M.\ M.\ Salomaa and G.\ E.\ Volovik, Phys.\ Rev.\ Lett.\ \textbf{56}, 363 (1986).
\bibitem{Fogelstroem1999}  M.\ Fogelström and J.\ Kurkijärvi, J.\ Low Temp.\ Phys.\ \textbf{116}, 1 (1999).
\bibitem{Salomaa1983} M.\ M.\ Salomaa and G.\ E.\ Volovik, Phys.\ Rev.\ Lett.\ \textbf{51}, 2040 (1983).
%\bibitem{footnote-helium3-b-phase-vortices} {
%Actually, this transition is, however, not observed in experiments \cite{Ikkala1982,Kondo1991}. This serious conflict between the theory and the experiments has not been solved yet and need to be clarified \cite{Fogelstroem1995}.
%}
%\bibitem{Ikkala1982} O.\ T.\ Ikkala, G.\ E.\ Volovik, P.\ J.\ Hakonen, Yu.\ M.\ Bun'kov, S.\ T.\ Islander, G.\ A.\ Kharadze, JETP Lett.\ \textbf{35}, 416 (1982).
%\bibitem{Kondo1991} Y.\ Kondo, J.\ S.\ Korhonen, M.\ Krusius, V.\ V.\ Dmitriev, Yu.\ M.\ Mukharsky, E.\ B. Sonin, G.\ E.\ Volovik, Phys.\ Rev.\ Lett.\ \textbf{67}, 81 (1991).


\bibitem{Smidman2017} M.\ Smidman, M.\ B.\ Salamon, H.\ Q.\ Yuan, and D.\ F.\ Agterberg, Rep.\ Prog.\ Phys.\ \textbf{80}, 036501 (2017).
\bibitem{Nagato1993} Y.\ Nagato, K.\ Nagai, and J.\ Hara, J.\ Low Temp.\ Phys.\ \textbf{93}, 33 (1993).
\bibitem{Schopohl1995} N.\ Schopohl and K.\ Maki, Phys.\ Rev.\ B \textbf{52}, 490 (1995).
\bibitem{Eschrig1999} M.\ Eschrig, J.\ A.\ Sauls, and D.\ Rainer, Phys.\ Rev.\ B \textbf{60}, 10447 (1999).
\bibitem{Nagai2012} Y.\ Nagai, K.\ Tanaka, and N.\ Hayashi, Phys.\ Rev.\ B \textbf{86}, 094526 (2012).
\bibitem{Shampine1986} L.\ F.\ Shampine, Math.\ Comp.\ \textbf{46}, 135 (1986).
\bibitem{Eyert1996} V.\ Eyert, J.\ Comput.\ Phys.\ \textbf{124}, 271 (1996).
\bibitem{Andreev1964} A.\ F.\ Andreev, Sov.\ Phys.\ JETP \textbf{19}, 1228 (1964).
\bibitem{Kulik1970} I.\ O.\ Kulik, Sov.\ Phys.\ JETP \textbf{30}, 944 (1970). 
\bibitem{Stone1996} M.\ Stone, Phys.\ Rev.\ B \textbf{54}, 13222 (1996).
\bibitem{Ichioka1996} M.\ Ichioka, N.\ Hayashi, N.\ Enomoto, and K.\ Machida, J.\ Phys.\ Soc.\ Jpn.\ \textbf{53}, 15316 (1996).
\bibitem{Kopnin2001} N.\ B.\ Kopnin, \textit{Theory of Nonequilibrium Superconductivity} (Oxford University Press, Oxford, 2001).
%\bibitem{Kita2001} T.\ Kita, Phys.\ Rev.\ B \textbf{64}, 054503 (2001).
%\bibitem{Hillier2012} A.\ D.\ Hillier, J.\ Quintanilla, B.\ Mazidian, J.\ F.\ Annett, R.\ Cywinski, Phys.\ Rev.\ Lett.\ \textbf{109} (2012).


\end{thebibliography}
\end{document}